\def\etal{{\frenchspacing\it et al.}}

\def\beq#1{\begin{equation}\label{#1}}
\def\eeq{\end{equation}}
\def\beqa#1{\begin{eqnarray}\label{#1}}
\def\eeqa{\end{eqnarray}}

\def\etal{{\frenchspacing\it et al.}}

\def\fun#1#2{\lower3.6pt\vbox{\baselineskip0pt\lineskip.9pt
        \ialign{$\mathsurround=0pt#1\hfill##\hfil$\crcr#2\crcr\sim\crcr}}}
        
\documentclass[12pt,preprint]{aastex}

\usepackage{graphicx}
\usepackage{epsfig}

\newcommand{\be}{\begin{equation}}
\newcommand{\ee}{\end{equation}}
\newcommand{\ba}{\begin{eqnarray}}
\newcommand{\ea}{\end{eqnarray}}

\shorttitle{Photometric Redshift Estimator for SNe~Ia}
\shortauthors{Wang et al. }

\begin{document}

\title{Survey Requirements for Accurate and Precise Photometric 
Redshifts for Type~Ia Supernovae}
\author{Yun~Wang\altaffilmark{1},
Gautham Narayan\altaffilmark{2}, and
Michael Wood-Vasey\altaffilmark{2}}
\altaffiltext{1}{Homer L. Dodge Department of Physics \& Astronomy, 
		 Univ. of Oklahoma,
                 440 W Brooks St., Norman, OK 73019;
                 email: wang@nhn.ou.edu}
\altaffiltext{2}{Center for Astrophysics, 
                 Harvard University, 
                 60 Garden Street, Cambridge, MA 02138; 
                 email: gnarayan@cfa.harvard.edu, wmwood-vasey@cfa.harvard.edu}
\begin{abstract}

In this paper we advance the simple analytic photometric redshift
estimator for Type~Ia supernovae (SNe~Ia) proposed by \citet{Wang07},
and use it to study simulated SN~Ia data.
We find that better than 0.5\% accuracy in $z_{\rm phot}$ 
(with $\sigma[(z_{\rm phot}-z_{\rm spec})/(1+z_{\rm spec})]<0.005$) is possible for 
SNe~Ia with well sampled lightcurves in three observed passbands ($riz$)
with a signal-to-noise ratio of 25 at peak brightness, if the extinction
by dust is negligible. The corresponding bias in $z_{\rm phot}$ 
(the mean of $z_{\rm phot}-z_{\rm spec}$) is $5.4\times 10^{-4}$.
If dust extinction is taken into consideration in the $riz$
observer-frame lightcurves, the accuracy in $z_{\rm phot}$ deteriorates to 4.4\%,
with a bias in $z_{\rm phot}$ of $8.0\times 10^{-3}$.
Adding the $g$ band lightcurve improves the
accuracy in $z_{\rm phot}$ to 2.5\%,
and reduces the bias in $z_{\rm phot}$ to $-1.5\times 10^{-3}$.
Our results have significant implications for the design
of future photometric surveys of SNe~Ia from both ground and space
telescopes. Accurate and precise photometric redshifts boost
the cosmological utility of such surveys.

\end{abstract}



\section{Introduction}

The discovery of cosmic acceleration \citep{Riess98,Perl99} was made using 
Type~Ia supernovae (SNe~Ia) as cosmological standard candles
\citep{Phillips93,Riess95}.
The unknown reason for the observed cosmic acceleration has
been dubbed ``dark energy''. 
Obtaining the spectroscopic redshifts of SNe~Ia is the most costly 
aspect of supernova surveys.
Large photometric surveys of SNe~Ia
have been planned to help illuminate the nature of dark energy.
Obtaining accurate and precise photometric redshifts (photo-$z$'s) 
will be key in maximizing the cosmological utility of such surveys \citep{Huterer04}.

In this paper, we advance the simple analytic photometric redshift
estimator for Type~Ia supernovae (SNe~Ia) proposed by \citet{Wang07}\footnote{Based
on the analytic photometric redshift estimator for galaxies proposed
by \cite{Wang98a}.},
and use it to study simulated SN~Ia data.
In order to derive unbiased and accurate photo-$z$'s,
it will be important to develop a variety of different techniques 
to cross-check sensitivities to various systematic uncertainties
and correctly calculating the covariance between distance modulus
and photo-$z$ when fitting for cosmological parameters.

We present our method in Sec.~\ref{sec:method} and the results in Sec.~\ref{sec:results}.
We discuss and summarize in Sec.~\ref{sec:discussion}.

\section{The Method}
\label{sec:method}

\subsection{The analytic photo-$z$ estimator}

The analytic photo-$z$ estimator for SNe~Ia proposed by \citet{Wang07}
is empirical, model independent (no templates used), 
and uses observables that reflect the properties
of SNe~Ia as calibrated standard candles.
It was developed using the SN~Ia data released by the Supernova
Legacy Survey \citep{Astier06}.

This estimator uses the fluxes in $griz$ (or $riz$) at the epoch 
of $i$ maximum flux to make an effective K-correction
to the $i$ flux. The first estimate of redshift is given by
\be
z_{\rm phot}^{0}=c_1 + c_2 g_f +c_3 r_f + c_4 i_f + c_5 z_f +c_6 i_f^2
+ c_7 i_f^3
\label{eq:z0}
\ee
where $g_f=2.5\log(f_g)$, $r_f=2.5\log(f_r)$, $i_f=2.5\log(f_i)$,
and $z_f=2.5\log(f_z)$, and $f_g$, $f_r$, $f_i$, $f_z$ are fluxes
in counts, normalized to some fiducial zeropoint, in $griz$ at the epoch of $i$ maximum flux.

Next, it calibrates each SN~Ia in its estimated rest-frame using
\be
\Delta i_{15}= 2.5 \log(f_i^{15d}/f_i),
\label{eq:del_i15}
\ee
where $f_i^{15d}$ is the $i$ band flux at 15 days after
the $i$ flux maximum in the estimated rest-frame, corresponding
to the epoch of $\Delta t^{15d}=15 (1+z_{\rm phot}^{0})$ days after the
epoch of $i$ flux maximum.	    

The final estimate for the photometric redshift is given by
\be
z_{\rm phot} = \sum_{i=1}^8 c_i \,p_i,
\label{eq:z_a}
\ee
where the data vector ${\mbox {\bf p}}=\{1, g_f, r_f, i_f, z_f, i_f^2,
i_f^3, \Delta i_{15}\}$.
The coefficients $c_i$ (i=1,2,...,8) are found by using
a training set of SNe~Ia with $griz$ (or $riz$, for which $c_2=0$) 
lightcurves and
measured spectroscopic redshifts. 
We use the jackknife technique \citep{Lupton93} to estimate the 
mean and the covariance matrix of $c_i$ (see Sec.~\ref{sec:results}).

\subsection{Simulation of data}

SN~Ia lightcurves for training and testing were simulated using 
the templates and various routines from MLCS2k2 \citep{Jha07}.  
We created two sets of $griz$ lightcurves using the MLCS2k2 ``final'' 
templates at random redshifts in the range $0.001 \leq z \leq 1$ 
with S/N=25 at peak and $\Delta$, 
the parameter that MLCS2k2 uses to describe the variation in the 
SN~Ia lightcurves, fixed to $0$. The luminosity distance $\mu$, 
is calculated from the redshift assuming
a flat cosmology with $\Omega_M=0.3$, $h=0.65$ and $w=-1$. The particular choice 
of cosmology will not affect the photo-$z$ estimator as long as the training and testing 
lightcurves are consistent with each other. 

The first training set consists of 100 lightcurves and was 
simulated with no host extinction. 
The corresponding blind test set contains 1000 lightcurves 
similar to those in the training set.

The second training set consists of 200 lightcurves, and has $A_V$ 
drawn at random from the ``default'' distribution, an exponential 
decay with argument $0.46$~mag and zero for $A_V < 0$ mag, following \citet{Jha07}. 
The particular form of this exponential tail is derived 
from the observation \citep{Lira95} that SNe~Ia have a common 
color at $\sim40$ days past peak. We assume the extinction 
law to be $R_V = 3.1$.  We used a larger training 
set of 200 simulated lightcurves for training to compensate 
for the larger variation in the data 
caused by including extinction. 
The corresponding blind test set contains 1000 lightcurves 
similar to those in the training set.

The actual training sets and blind test sets used are
slightly smaller than described above, because
only SNe~Ia with $riz$ lightcurves, i.e. $z\le0.95$, are used in this paper.
We use the training sets to derive the photo-$z$ coefficients
and their covariance matrix (see Eq.~\ref{eq:z_a} and Sec.~\ref{sec:results}), 
and the blind test sets to evaluate the performance of the 
photo-$z$ estimator.   

\section{Results}
\label{sec:results}

We use a slightly modified version of the jackknife technique 
\citep{Lupton93} to estimate the 
mean and the covariance matrix of $c_i$. 
From the training set containing $N$ SNe~Ia, 
we extract $N$ subsamples each containing $N-1$ SNe~Ia
by omitting one SN~Ia. The coefficients $c_i^{(s)}$ (i=1,2,...,8)
for the $s$-th subsample are found by a maximum likelihood
analysis matching the predictions of
Eq.~\ref{eq:z_a} with the spectroscopic redshifts.

The mean of the coefficients $c_i$ (i=1,2,...,8) are given by
\be
\langle c_i \rangle = \frac{1}{N} \sum_{s=1}^{N} c_i^{(s)}.
\label{eq:mean}
\ee
Note that this is related to the usual ``bias-corrected jackknife
estimate'' for $c_i$, $c_i^J$, as follows:
\be
c_i^J \equiv c_i^N +(N-1) \left(c_i^N-\langle c_i \rangle\right),
\ee
where $c_i^N$ are estimated from the entire training set (with $N$ SNe~Ia).
We found that for small training sets (with $N<20$) that include
SNe~Ia at $z \sim 0$, $c_i^J$ give
biased estimates of $c_i$ by giving too much weight to the
SN~Ia with the smallest redshift. For training sets 
not including nearby SNe~Ia, $\langle c_i \rangle$ and $c_i^J$ 
are approximately equal.
We have chosen to use $\langle c_i \rangle$ from Eq.~\ref{eq:mean}
as the mean estimates for $c_i$ to avoid biased $z_{\rm phot}$ 
for SNe~Ia at $z$ close to zero.

The covariance matrix of $c_i$ (i=1,2,...,8) are given by
\be
{\rm Cov}(c_i,c_j)= \frac{N-1}{N} \sum_{s=1}^{N} 
\left(c_i^{(s)} - \langle c_i \rangle \right)\, 
\left(c_j^{(s)} - \langle c_j \rangle \right)
\ee
Since $z_{\rm phot}= \sum c_i p_i$, $\Delta z_{\rm phot}= \sum p_i\Delta c_i$,
since the uncertainty in $z_{\rm phot}$ is dominated by the
uncertainty in $c_i$.
Therefore estimated error in $z_{\rm phot}$ is
\be
{\rm d}z_{\rm phot}= \left\{ \sum_{i=1}^8 \sum_{j=1}^8 
 [p_i] \, {\rm Cov}(c_i,c_j) [p_j] \right\}^{1/2},
\ee
where ${p_i}$ and ${p_j}$ (i,j=1,2,...8) are
the $i$-th and $j$-th components of the data vector
${\mbox {\bf p}}=\{1, g_f, r_f, i_f, z_f, i_f^2,
i_f^3, \Delta i_{15}\}$. 

In applying the photo-$z$ estimator to simulated data, we find
that applying it to the entire data set leads to substructure
in the estimated $z_{\rm phot}$. Therefore, we divide the
training set into three $i$ maximum flux ranges (corresponding 
to three roughly equal redshift intervals), 
and applying the photo-$z$ estimator to each flux range.

Fig.~\ref{fig1} shows the resultant performance of
the photo-$z$ estimator for simulated SN~Ia lightcurves
in $riz$, with S/N=25 at peak brightness, and zero
extinction due to dust. The top panel shows the
training set of 97 SNe~Ia, with different point types and
color denoting different flux ranges.
The $c_i$'s derived from the training set are used
to predict $z_{\rm phot}$ for a blind-test set of 940 SNe~Ia
(divided into the same flux ranges as the training set),
with the results shown in the bottom panel of Fig.~\ref{fig1}.

The results for the blind-test set closely mimic that
of the training set.
Since the data are very sparse at $z<0.02$ due to
the small size of the training set, 
the estimated uncertainty on $z_{\rm phot}$, d$z_{\rm phot}$,
becomes large for nearby SNe~Ia.
Since d$z_{\rm phot}$ is a reliable
indicator of the accuracy of $z_{\rm phot}$,
we can use d$z_{\rm phot}$ to exclude the SNe~Ia
with estimated $z_{\rm phot}$ not suitable for inclusion
in the cosmological data set.

We find that d$z_{\rm phot}<0.1$ only excludes 9 out of the
940 SNe~Ia in the blind-test set, all at $z<0.02$.
For the culled blind-test set of 931 SNe~Ia (d$z_{\rm phot}<0.1$),
$\sigma[(z_{\rm phot}-z_{\rm spec})/(1+z_{\rm spec})]=0.0043$.
The corresponding bias in $z_{\rm phot}$ 
(the mean of $z_{\rm phot}-z_{\rm spec}$) is $5.4\times 10^{-4}$ overall
and is  $[0.48,  0.79,  1.5, -0.81]\times 10^{-3}$ in 
the 4 redshift bins bounded by $[0,0.25,0.5,0.75,1]$.

Fig.~\ref{fig2} shows the performance of
the photo-$z$ estimator for simulated SN~Ia lightcurves
in $riz$, with S/N=25 at peak brightness, and 
extinction due to dust parameterized by $A_V$.
Again, we cull the blind-test set by requiring
that d$z_{\rm phot}<0.1$; this excludes 16 out of the
956 SNe~Ia, most of which are highly extinguished or
have very low $z$.
The inclusion of dust extinction leads to a deteriorated
accuracy in $z_{\rm phot}$ of 4.4\%.
The corresponding bias in $z_{\rm phot}$ 
is $8.0\times 10^{-3}$ overall
and is  $[-9.3, -22.,  25.,  53.]\times 10^{-3}$ in the 
4 redshift bins bounded by $[0,0.25,0.5,0.75,1]$.

Fig.~\ref{fig3} is similar to Fig.~\ref{fig2}, but only the subsets of data
(both the training set and the blind-test set) with 
$g$ band lightcurves are included.
Note that the $i$ maximum flux ranges in Fig.~\ref{fig3} are similar
to Fig.~\ref{fig2}, but slightly adjusted to make sure that
the sub sample with the smallest fluxes contains
enough SNe~Ia. The SNe~Ia at $z>0.54$ in
our simulated data set do not have $g$ band lightcurves,
as these require rest-frame lightcurve templates bluer than $u$ band,
which are not yet readily available.

Imposing d$z_{\rm phot}<0.1$ excludes 12 out of the
537 SNe~Ia, most of which are highly extinguished or
have very low $z$.
Adding the $g$ band lightcurves improve the
accuracy in $z_{\rm phot}$ to 2.5\%.
The corresponding bias in $z_{\rm phot}$ 
is reduced to $-1.5\times 10^{-3}$ overall
and is $[-6.0, 3.2]\times 10^{-3}$ in the 
2 redshift bins bounded by $[0,0.266,0.532]$.

\section{Discussion and Summary}
\label{sec:discussion}

Accurate and precise photo-$z$'s will be critical for 
enabling cosmology with future large SNe~Ia photometric surveys
 \citep{Huterer04}.
In this paper, we have advanced the empirical and analytic photo-$z$
estimator proposed by \citet{Wang07} and used it
to study simulated data in order to derive
the survey requirements for accurate and precise photo-$z$'s
for SNe~Ia from large photometric surveys.
Future work will explore the critically-important question
of the degree of covariance between the redshift determined 
using this method and the distance modulus
determined by using a luminosity distance fitter
since it is these two parameters that form the basis
for the cosmological measurements with SNe~Ia.

We apply the photo-$z$ estimator to $riz$ data divided into
$i$ maximum flux ranges that roughly correspond to
equal redshift intervals (see Fig.~\ref{fig1} and Fig.~\ref{fig2}). 
We use d$z_{\rm phot}<0.1$ as a quality measurement to cull the blind-test set.
This cut only excludes $<2$\% of the SNe~Ia,
most of which are highly extinguished or are
at very low $z$. 
To derive accurate cosmological constraints,
it is optimal to combine the very large set
of SNe~Ia with well estimated photo-$z$'s with
a small spectroscopic sample of SNe~Ia at very low $z$ 
expected from ongoing nearby surveys
\citep{Huterer04}. 
SNe~Ia at $z<0.025$ are currently excluded
from SN~Ia cosmological analyses even for spectroscopic
data, because peculiar velocities due to cosmic large
scale structure modifies the Hubble diagram of
nearby standard candles \citep{Wang98b}, and
have the largest impact on the nearest SNe~Ia (see, e.g., 
\cite{Zehavi98,Cooray06,Hui06,Conley07}).

We find that determining $z_{\rm phot}$'s with better than 0.5\% accuracy 
(RMS[$\Delta z_{\rm phot}/(1+z_{\rm spec})]$ = 0.0043\%)
and negligible bias ($\langle z_{\rm phot}-z_{\rm spec}\rangle=5.4\times 10^{-4}$)
is possible if dust extinction can be neglected, for
SNe~Ia with well sampled lightcurves in $riz$ and S/N=25
at peak brightness.
The inclusion of dust extinction leads to a drastic deterioration
of the accuracy of $z_{\rm phot}$
(RMS[$\Delta z_{\rm phot}/(1+z_{\rm spec})]$ = 4.4\%,
and $\langle z_{\rm phot}-z_{\rm spec}\rangle =8.0\times 10^{-3}$).
Adding the $g$ band lightcurve significantly improves
the accuracy and precision of $z_{\rm phot}$
with RMS[$\Delta z_{\rm phot}/(1+z_{\rm spec})$] = 2.5\%,
and $\langle z_{\rm phot}-z_{\rm spec}\rangle = -1.5\times 10^{-3}$.

Future supernova surveys can easily obtain 
the multi-band photometry of a huge number of supernovae
\citep{Wang00,JEDI,Phillips06}.
We only considered $griz$ photometry in this paper,
due to the limitation of currently available SN Ia lightcurve
templates. The addition of photometry at longer wavelengths
should provide more dramatic improvement on the
accuracy and precision of $z_{\rm phot}$ since extinction
decreases with wavelength, and SNe~Ia are better standard
candles in the near IR \citep{KK04}.
In future work, we will study simulated SN~Ia data
over the entire wavelength range of relevance to future
observations, and examine the impact of realistic survey 
and observational parameters on the precision and
accuracy of our photo-$z$ estimator.

Our results are encouraging for the prospect of using photometric
surveys to constrain cosmology, such as those planned for 
the Advanced Liquid-mirror Probe for Astrophysics, Cosmology and 
Asteroids (ALPACA)\footnote{http://www.astro.ubc.ca/LMT/alpaca/};
Pan-STARRS\footnote{http://pan-starrs.ifa.hawaii.edu/}; 
the Dark Energy Survey (DES)\footnote{http://www.darkenergysurvey.org/};
and the Large Synoptic Survey Telescope (LSST)\footnote{http://www.lsst.org/}.

\bigskip

{\bf Acknowledgments}
We thank the Aspen Center for Physics for hospitality 
where part of the work was done.
We are grateful to Alex Kim for useful discussions.
YW  is supported in part by NSF CAREER grant AST-0094335. 
MWV is supported in part by NSF AST-057475.

\clearpage
\setcounter{figure}{0}

\begin{figure}
\includegraphics{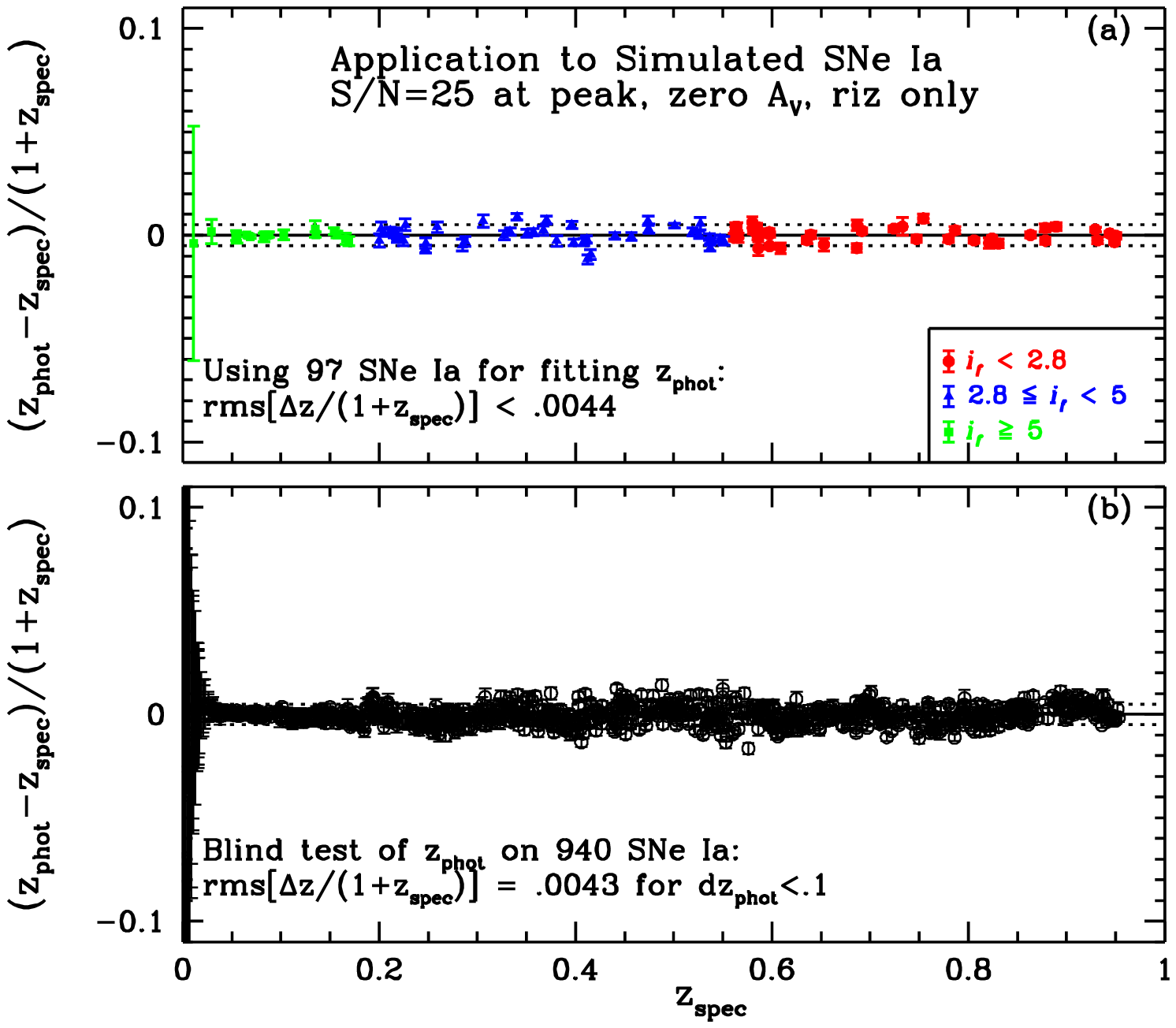}
\caption{The analytic photo-$z$ estimator 
applied to simulated SN~Ia lightcurves
in $riz$, with S/N=25 at peak brightness, and zero
extinction due to dust. 
}
\label{fig1}
\end{figure}

\begin{figure}

\includegraphics{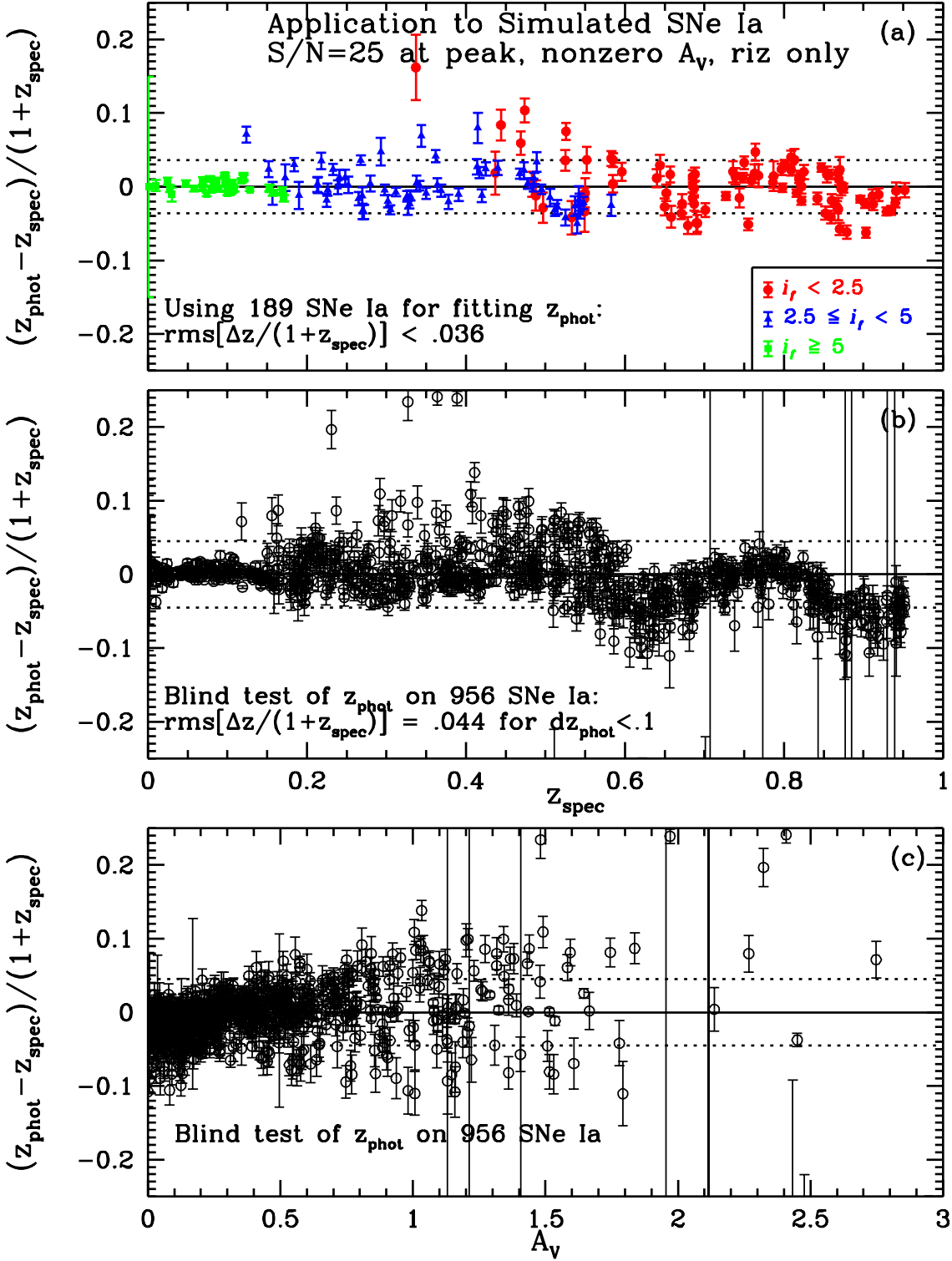}
\caption{The analytic photo-$z$ estimator 
applied to simulated SN~Ia lightcurves
in $riz$, with S/N=25 at peak brightness, and 
dust extinction parameterized by $A_V$. 
Dust extinction leads to deterioration in
the accuracy of $z_{\rm phot}$.
The y-axis range is
the same as in Fig.~\ref{fig3}.
}
\label{fig2}
\end{figure}

\begin{figure}
\includegraphics{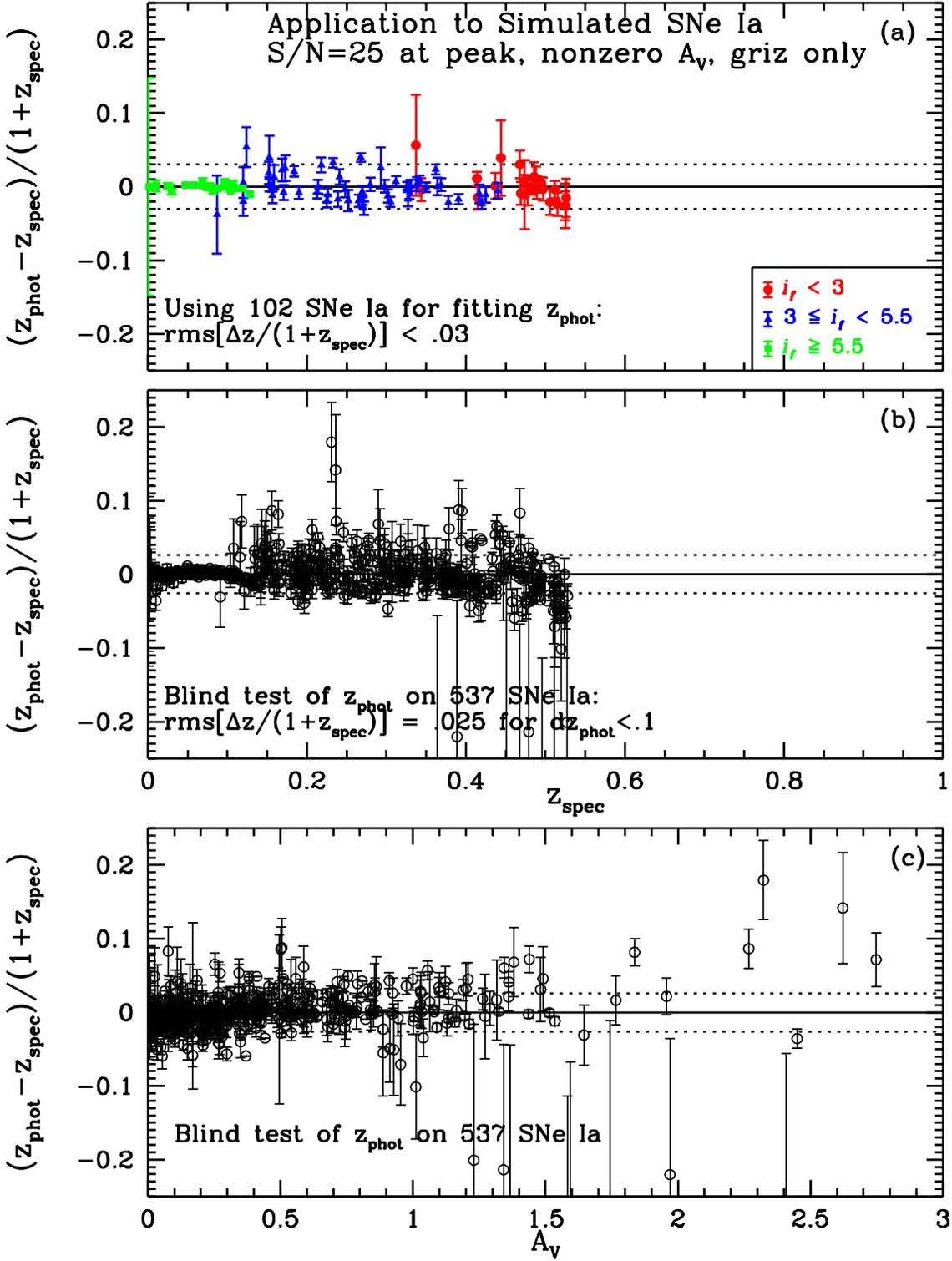}
\caption{The analytic photo-$z$ estimator 
applied to simulated SN~Ia lightcurves
in $griz$, with S/N=25 at peak brightness, and 
dust extinction parameterized by $A_V$. 
Adding the $g$ band lightcurves improves the
accuracy of $z_{\rm phot}$.
The y-axis range is
the same as in Fig.~\ref{fig2}.
}
\label{fig3}
\end{figure}

\end{document}